\documentclass[fleqn,10pt]{article} 
\usepackage[left=2cm,%
                right=2cm,%
                top=2.3cm,%
                bottom=2.3cm,%
                headheight=12pt]{geometry}%

\usepackage[noadjust]{cite}
\usepackage[colorlinks=true, allcolors=blue]{hyperref}

\usepackage[utf8]{inputenc}
\usepackage[english]{babel}
\usepackage[T1]{fontenc}

\usepackage{times}      
\usepackage{mathptmx}   

\usepackage{amsmath,amsfonts,amssymb}
\usepackage{graphicx}

\usepackage[title]{appendix}		
\usepackage{breqn}		
\usepackage{braket}          
\allowdisplaybreaks
\usepackage[labelfont={bf,sf},%
                labelsep=period,%
                justification=justified]{caption}

\usepackage{authblk}
\usepackage{fancyhdr}  
\usepackage{lastpage}  
\usepackage[explicit]{titlesec}
\titleformat{name=\section,numberless}
  {\large\sffamily\bfseries}
  {}
  {0em}
  {#1}
  []  
\titleformat{\subsection}
  {\sffamily\bfseries}
  {\thesubsection}
  {0.5em}
  {#1}
  []
\titleformat{\subsubsection}
  {\sffamily\small\bfseries\itshape}
  {\thesubsubsection}
  {0.5em}
  {#1}
  []    
\titleformat{\paragraph}[runin]
  {\sffamily\small\bfseries}
  {}
  {0em}
  {#1} 
\titlespacing*{\section}{0pc}{3ex \@plus4pt \@minus3pt}{5pt}
\titlespacing*{\subsection}{0pc}{2.5ex \@plus3pt \@minus2pt}{0pt}
\titlespacing*{\subsubsection}{0pc}{2ex \@plus2.5pt \@minus1.5pt}{0pt}
\titlespacing*{\paragraph}{0pc}{1.5ex \@plus2pt \@minus1pt}{10pt}

\makeatletter
\renewcommand{\@maketitle}{%
{\raggedright\sffamily\bfseries\fontsize{20}{25}\selectfont \@title\par}%
\vskip10pt
{\raggedright\sffamily\fontsize{12}{16}\selectfont \@author\par}
}%

\let\oldbibliography\thebibliography
\renewcommand{\thebibliography}[1]{%
\oldbibliography{#1}%
\setlength\itemsep{0pt}%
}

\newcommand{\commas}[1]{``{#1}''}
\newenvironment{eqnlist}{\subequations\align}{\endalign\endsubequations}

\title{A continued fraction based approach for the Two-photon Quantum Rabi Model}
\author[1,2,*]{Elena Lupo}
\author[1,3]{Anna Napoli}
\author[3,4]{Antonino Messina}
\author[5,6,7]{Enrique Solano}
\author[8]{\'{I}\~{n}igo L. Egusquiza}

\affil[1]{Department of Physics and Chemistry, University of Palermo, Via Archirafi 36, I-90123 Palermo, Italy}
\affil[2]{Advanced Technology Institute and Department of Physics, University of Surrey, Guildford, GU2 7XH, UK}
\affil[3]{I.N.F.N. Sezione di Catania}
\affil[4]{Department of Mathematics and Computer Science, University of Palermo, Via Archirafi 34, I-90123 Palermo, Italy}
\affil[5]{Department of Physical Chemistry, University of the Basque Country
UPV/EHU, Apartado 644, 48080 Bilbao, Spain}
\affil[6]{IKERBASQUE, Basque Foundation for Science, Maria Diaz de Haro 3, 48013
Bilbao, Spain}
\affil[7]{Department of Physics, Shanghai University, 200444 Shanghai, China}
\affil[8]{Department of Theoretical Physics and History of Science, University of the Basque Country UPV/EHU, Apartado 644, 48080 Bilbao, Spain}

\affil[*]{e.lupo@surrey.ac.uk}

\begin{document}
\maketitle
\vspace{5pt}
\section*{Abstract}
\vspace{5pt}
We study the Two Photon Quantum Rabi Model by way of its spectral functions and survival probabilities. This approach allows numerical precision with large truncation numbers, and thus exploration of the spectral collapse. We provide independent checks and calibration of the numerical results by studying an exactly solvable case and comparing the essential qualitative structure of the spectral functions. We stress that the large time limit of the survival probability provides us with an indicator of spectral collapse, and propose a technique for the detection of this signal in the current and upcoming quantum simulations of the model.

\thispagestyle{empty}
\vspace{10pt}
\section*{Introduction}
\label{Intro}
The Quantum Rabi Model (QRM) and the Two-Photon Quantum Rabi Model (2$\gamma$QRM) represent two basic models for the description of the interaction of light and matter. The first one describes a two-level system bilinearly coupled to  a quantized bosonic field mode; 2$\gamma$QRM is one of  its simplest  generalizations, in which the interaction term is now quadratic in the annihilation and creation bosonic operators. The bilinear QRM  for light-matter interaction appeared  more than 80 years ago \cite{Rabi, Rabi2, JC}. Yet  interest in this model has never waned and, rather,  it has even grown recently. This growth  mainly stems from its potential application to platforms used for quantum technologies \cite{NoriAt}. The QRM depends on two independent parameters, and the  dynamical properties of the atom mode  are qualitatively very different in different regions of the parameter space. Most of the experimental Cavity Quantum Electrodynamics (CQED) setups are characterized by physical conditions inside the  weak-coupling regime, in which  the Quantum Rabi model can be effectively simplified to the exactly treatable Jaynes-Cummings model. So as to best describe new, more advanced quantum devices, such as  superconducting  circuits or trapped ions systems, for instance, the description of the  QRM must be extended to the  appropriate regions of the parameter space, for which the Jaynes-Cummings approximation fails.\\
Alternatively, one can view these newer platforms as `Quantum Simulators' \cite{NoriQS, SCQS, PedernalesIons}, in which one can  realize models that had been previously discarded as `unphysical'.  In fact, coupling constant values much higher than the ones typical of CQED setups have been  measured in the last years, even reaching the so-called Ultrastrong Coupling (USC, $0.1\,\omega\ll g \ll \omega$) and the Deep Strong Coupling (DSC, $g\gg \omega$) regimes in the context of circuit Quantum Electrodynamics cQED \cite{Exper1, Exper2, Exper3}.\\
In this vein of Quantum Simulation, other possibilities have appeared. For instance, the interaction Hamiltonian for trapped ions is non-linear, thus allowing  this system to be exploited in order to  investigate the dynamics of various QRM generalizations \cite{PedernalesIons, Felicetti}. For these reasons, the interest in the QRM and its variants has been rekindled, and a strong effort  to construct their solutions and to clarify the relative dynamical properties is under way \cite{SolanoDSC, SolanoSpec, Schweber, Braak, BraakModels, Bogoliubov, XieRev}. A major role in these new developments has been played  by the  analytic solutions of the QRM, found first in 2011 \cite{Braak} and based on its representation in the Bargmann space of the holomorphic functions  \cite{Bargmann}. Other approaches, exploiting a suitable Bogoliubov transformation \cite{Bogoliubov}, or an expansion in the basis of  Heun functions, have also been proposed \cite{Heun1, Heun2}.\\
Among the many generalizations of the QRM, the Two-photon Quantum Rabi Model (2$\gamma$QRM)  is of particular interest. It was introduced as an effective model for a three-level system interacting with a bosonic mode in which the intermediate level can be adiabatically eliminated \cite{Sukumar, Davidovich, Bullough, Toor}. Even if the original phenomenological model was treated in the Rotating Wave Approximation \cite{Sukumar}, some work on the influence of the counter-rotating terms has been carried out in the past \cite{PengLi, NgLoLiu, Bishop, Albert}. The  more recent possibility of
realizing the 2$\gamma$QRM in quantum simulators has sparked a new flurry of studies. In particular one should notice the recent proposal for  its implementation in trapped ions systems and superconducting circuits  \cite{Felicetti,PhysRevA.95.063844,SolanoSimSC,Schneeweiss:2017aa}. Moreover, after  Braak's solution for the QRM, the same approach was applied to the determination of the 2$\gamma$QRM spectrum using  G-functions \cite{Travenec, Travenec2}. Alternatively the search of the analytic solution has also been presented as an expansion  in the generalized squeezed number states \cite{Bogoliubov, Zhang, Zhang2}. 
An important feature of the model, namely  the collapse of the discrete spectrum into a continuum at a value of the coupling constant  $g=\omega/2$ \cite{Duan}, has also been made evident both with squeezed states  \cite{Metha} and with the Bargmann space \cite{Bargmann} approach.\\
Nonetheless,   useful as these analytical approaches are for  the spectrum as a function of the coupling strength between the fermionic and bosonic  subsystems, an  analytical form of the eigenstates, and thus  of all quantities of interest, is still to be obtained \cite{XieRev}. One such quantity, of particular relevance from an experimental point of view, is the spectral function, defined as  $\rho(E,\ket{\Psi})=\braket{\Psi|\delta(E-H)|\Psi}$, in terms of a  generic state $\ket{\Psi}$ of the system. In fact $\rho(E,\ket{\Psi})$ contains all the information useful for generating the time evolution of $\ket{\Psi}$, namely  those eigenvalues of the Hamiltonian whose eigenfunctions have an overlap with $\ket{\Psi}$  and the relative transition probabilities.\\
In this paper we put forward the spectral analysis of a factorized state  $\ket{n,\sigma}\equiv \ket{n}\ket{\sigma}$ of the 2$\gamma$QRM, $n$ being the eigenvalue of the number operator $a^\dagger a$ and $\sigma$ being the eigenvalue of the spin operator $\sigma_z$. This approach, valid in each point of the parameter space, is an alternative to the  Bargman solution of the model. To achieve this goal the relevant matrix element of the resolvent is presented in continued fraction form. We have thus direct access to two complementary quantities of interest: the spectral density and the survival probability. The structure of our approach allows us clean access to the  dynamics of the system near the collapse point of the 2$\gamma$QRM corresponding to the value of $g=0.5 \omega$.\\
The paper is organized as follows: in section \ref{sec2} we present the model, and apply a unitary transformation such that the eigenstates factorize in a bosonic and a spin part \cite{Messina}; in section \ref{sec3} we exploit the connection between the resovent of a tridiagonal matrix and  continued fractions to obtain a numerical determination for the spectral function of factorized states $\ket{n,\sigma}$; finally, in section \ref{sec4} we use the previous results to study the survival probability of the vacuum state of the system. 

\section{The two-photon model} \label{sec2}
The Two-photon Quantum Rabi Model (2$\gamma$QRM) presents an interaction which is  non linear  in the bosonic operators. The Hamiltonian can be expressed as: 
\begin{equation} \label{1}
H= \omega a^\dagger a + \frac{\omega_0}{2} \sigma_z + g (a^2 + (a^\dagger)^2) \sigma_x
\end{equation}
where $\omega$ is the frequency of the bosonic mode, $\omega_0$ the atomic frequency and $g$ the coupling constant between the two subsystems. Here and subsequently we set $\hbar$ to 1. The spectrum of this model has been numerically calculated in many works \cite{NgLoLiu, Bogoliubov, Felicetti, Travenec, Travenec2, Zhang, Zhang2, Duan}, and a link with the squeezed number states has been  pointed out \cite{Bogoliubov, Zhang, Zhang2}. This link  provides us with a better understanding of the spectrum collapse at $g=\omega/2$  \cite{Felicetti, Duan}, as follows.
Define, as usual,  the squeezing operator $S(\beta)=e^{-\frac{\beta}{2}\big( a^2 - (a^\dagger)^2\big)}$, and consider the 2$\gamma$QRM for $\omega_0=0$, written in the basis for which $\sigma_x$ is diagonal. Under squeezing transformations with squeezing parameters  $\beta_\pm=\pm\frac{1}{2}\tanh^{-1}\big(\frac{2g}{\omega}\big)$ one of the diagonal elements of the Hamiltonian becomes a harmonic oscillator. Clearly. the limit $|g|\to \omega/2$ is the limit of infinite squeezing.  This entails,  in  what regards the spectrum, the collapse of eigenvalues into a continuum in the limit  $g\to0.5\omega$,
and the (generalized) eigenstates are no longer  normalizable\cite{Felicetti, Duan}. As in \cite{Felicetti}, one can rewrite the Hamiltonian (\ref{1}) in terms of the position and momentum operators of the oscillator, $x=\sqrt{\frac{1}{2\omega}}(a+a^\dagger)$ and $p=i\sqrt{\frac{\omega}{2}}(a^\dagger-a)$, with unit mass:
\begin{equation}
H=\frac{\omega}{2}\Big\{ (\omega-2g\sigma_x)\frac{p^2}{\omega^2} + (\omega+2g\sigma_x)x^2\Big\} + \frac{\omega_0}{2}\sigma_z - \frac{\omega}{2}
\end{equation}
For $g<\omega/2$  the effective potential makes the system stable, while at the point $g=\omega/2$ one of the two quantities $x^2$ or $p^2$ disappears and the spectrum collapses into a continuum. This is immediately obvious if $\omega_0=0$. Were this parameter different from zero, isolated eigenstates would appear. In the context of an analysis of the asymptotic behaviour of solutions in  Bargmann space, the collapse point coincides with the limit situation, for which the eigenfunction is no longer normalizable \cite{Felicetti, BraakModels, Duan}.\\ 
As is well known, the QRM Hamiltonian commutes with a parity operator, and its eigenvalues can be arranged in parity subspaces.  In the case of the 2$\gamma$QRM the symmetry is $\mathbb{Z}_4$, since the Hamiltonian commutes with $\Pi_4=-e^{i\frac{\pi}{2}a^\dagger a}\sigma_z$, whose eigenvalues are the quartic roots of unity $\left\{\pm1,\pm i\right\}$. It follows that the full Hilbert space is organized in four infinite-dimensional chains:
\begin{equation}
\label{3}
\begin{split}
\ket{0,-} \leftrightarrow \ket{2,+} \leftrightarrow \ket{4,-} \leftrightarrow \ket{6,+} \leftrightarrow \ket{8,-} \leftrightarrow \cdots
 \\
 \ket{1,+} \leftrightarrow \ket{3,-} \leftrightarrow \ket{5,+} \leftrightarrow \ket{7,-} \leftrightarrow \ket{9,+} \leftrightarrow \cdots
\\
\ket{0,+} \leftrightarrow \ket{2,-} \leftrightarrow \ket{4,+} \leftrightarrow \ket{6,-} \leftrightarrow \ket{8,+} \leftrightarrow \cdots
 \\
\ket{1,-} \leftrightarrow \ket{3,+} \leftrightarrow \ket{5,-} \leftrightarrow \ket{7,+} \leftrightarrow \ket{9,-} \leftrightarrow \cdots
\end{split}
\end{equation}
We denote the corresponding four infinite-dimensional subspaces $S_w$, with $w\in\left\{\pm1\,,\pm i\right\}$. For instance, the vacuum state $\ket{0,-}\equiv \ket{0}\ket{-1}$ belongs to the subspace $S_{+1}$. Explicitly,
\begin{eqnlist} 
&{} \ket{\Psi_w}= \sum_{n=0}^{\infty} a_n \ket{2n, \sigma=-w\cos(\pi n)}, \qquad \textrm{for}\ w=\pm 1; \\
&{} \ket{\Psi_w}= \sum_{n=0}^{\infty} a_n \ket{2n+1, \sigma=-iw\cos(\pi n)}, \qquad \textrm{for}\ w=\pm i\,.
\end{eqnlist}

We shall now apply a transformation which factorizes the state $\ket{\Psi_w}$ into a bosonic and an atomic part, following the procedure of  \cite{Messina} for the QRM. In other words \cite{SolanoDSC}, we  use the parity basis. This factorization is indeed achieved with the  rotation
\begin{equation}
T=e^{-i\frac{\pi }{4}(\sigma_x -1)a^\dagger a}=\frac{1}{2}(1-\sigma_x)e^{i\frac{\pi}{2} a^\dagger a}+\frac{1}{2}(1+\sigma_x)\,.
\end{equation}
This rotation transfoms the Hamiltonian into $\tilde{H}= T^\dagger H T$, explicitly
\begin{equation}
  \label{eq:rotated}
\tilde{H} = \omega a^\dagger a + \frac{\omega _0}{2} \cos\Big(\frac{\pi}{2}a^\dagger a \Big)\sigma_z + \frac{\omega _0}{2} \sin\Big(\frac{\pi}{2}a^\dagger a \Big)\sigma_y + g ((a)^2 + (a^\dagger)^2 )\,. 
\end{equation}
The coupling term is now expressed as diagonal in the bosonic number operator. Under this rotation the subspaces of constant 4-parity become:
\begin{equation}
\label{subspaces}
\begin{split}
\tilde{S}_{-1} = \{ \ T^\dagger\ket{0,+},\ T^\dagger\ket{2,-},\ \cdots \ \} &{}\equiv \{ \ket{2n,+},\ n \in \mathbb{N}\ \}
\\
\tilde{S}_{+1} = \{ \ T^\dagger\ket{0,-},\ T^\dagger\ket{2,+},\ \cdots \ \} &{}\equiv \{ \ket{2n,-},\ n \in \mathbb{N}\ \}
\\
\tilde{S}_{-i} = \{ \ T^\dagger\ket{1,-},\ T^\dagger\ket{3,+},\ \cdots \ \} &{}\equiv \Big\{ \ket{2n+1}\otimes\frac{1}{\sqrt{2}}\big(\ket{+}+ i \ket{-}\big),\ n \in \mathbb{N}\  \Big\}
\\
\tilde{S}_{+i} = \{ \ T^\dagger\ket{1,+},\ T^\dagger\ket{3,-},\ \cdots \ \} &{}\equiv \Big\{ \ket{2n+1}\otimes\frac{1}{\sqrt{2}}\big(\ket{+}- i \ket{-}\big),\ n \in \mathbb{N}\  \Big\}
\end{split}
\end{equation}
and the Hamiltonian projected into each subspace is a quadratic of the bosonic creation and annihilation operators,
\begin{equation}
\label{9}
\begin{split}
\tilde{H}_{\pm 1} = &{}\omega a^\dagger a + g\big(a^2 + (a^\dagger)^2\big) \mp \frac{1}{2}\omega_0 \cos\Big(\frac{a^\dagger a}{2} \pi \Big) \\
 \tilde{H}_{\pm i} = &{}\omega a^\dagger a + g\big(a^2 + (a^\dagger)^2\big) \mp \frac{1}{2}\omega_0 \cos\Big(\frac{a^\dagger a-1}{2}\pi \Big)\,.
 \end{split}
 \end{equation}
 Thus
 each effective Hamiltonian $\tilde{H}_w$ is explicitly tridiagonal in the Fock basis.
 
\section{The Spectral Function and the Resolvent} \label{sec3}
\begin{figure}
\center
\includegraphics[height=6cm, width=14.16cm]{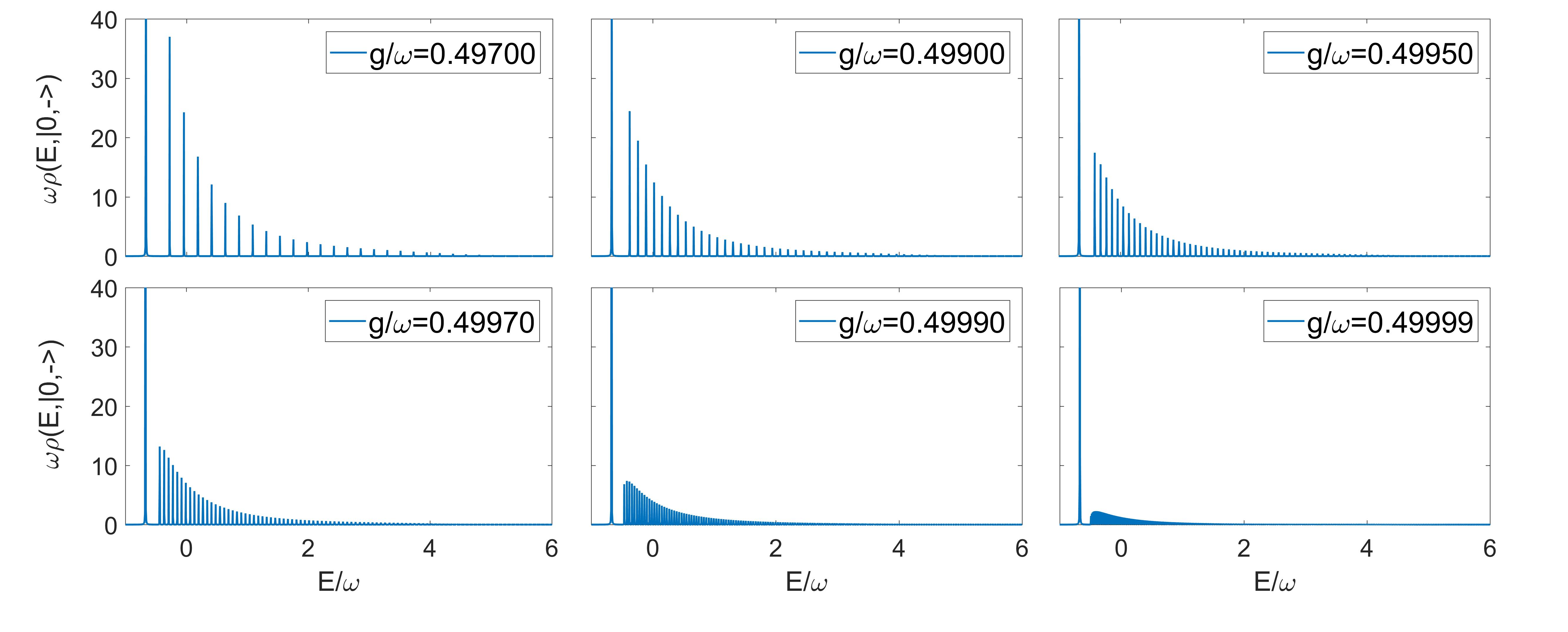}
\captionsetup{width=0.9\linewidth}
\caption{\label{fig:2}
Spectral density related  to the state $\ket{0,-}$ of the $2\gamma$QRM at different values of $g/\omega$, close to the collapse point $g/\omega=0.5$. The value of the atomic frequency is $\omega_0=0.8\omega$ in all cases, while $\varepsilon=0.0005$ (see eq. (\ref{ro})). The truncation of the continued fraction is chosen in order to have convergence.}
\end{figure}

\begin{figure}
\centering
\captionsetup{width=0.9\linewidth}
\includegraphics[height=6cm, width=14.16cm]{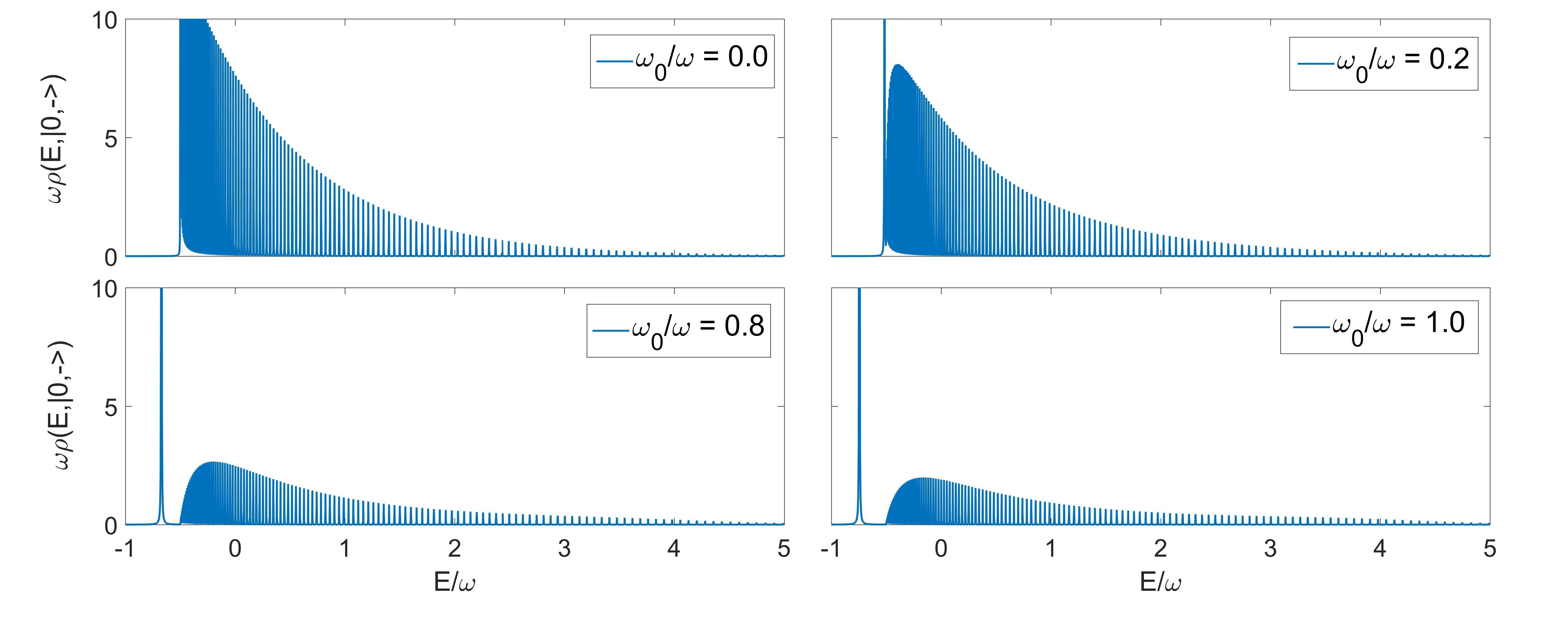}
\caption{\label{Om0a}Spectral function related to the vacuum state $\ket{0,-}$ of the 2$\gamma$QRM in correspondence of the collapse point $g/\omega=0.5$, at different values of $\omega_0$. It can be seen that the position of the isolated ground state is dependent on the value of the atomic frequency. In all cases $\varepsilon=0.0005$ (see eq. (\ref{ro})), while the truncation number exploited for the continued fraction is  $N=8000$.}
\end{figure}

\begin{figure}
\centering
\captionsetup{width=0.9\linewidth}
\includegraphics[clip,scale=0.07]{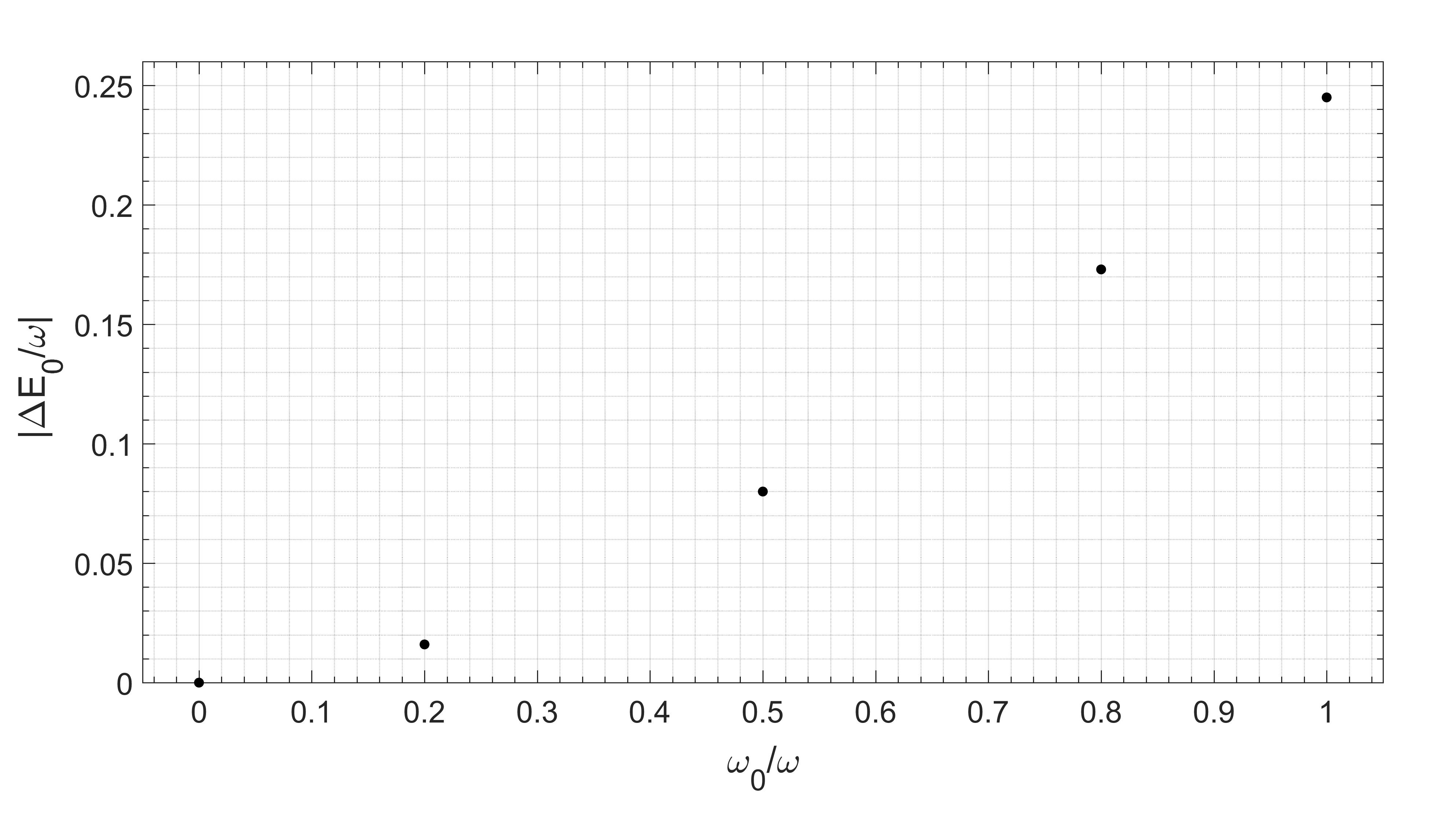}
\hspace*{2pt}\caption{\label{Om0b}Energy difference between the ground state and the continuum part of the spectrum in varying the two-level parameter $\omega_0$.}
\end{figure}

\begin{figure}[t]
\captionsetup{width=.95\linewidth}
\qquad\includegraphics[height=6.78cm, width=16cm]{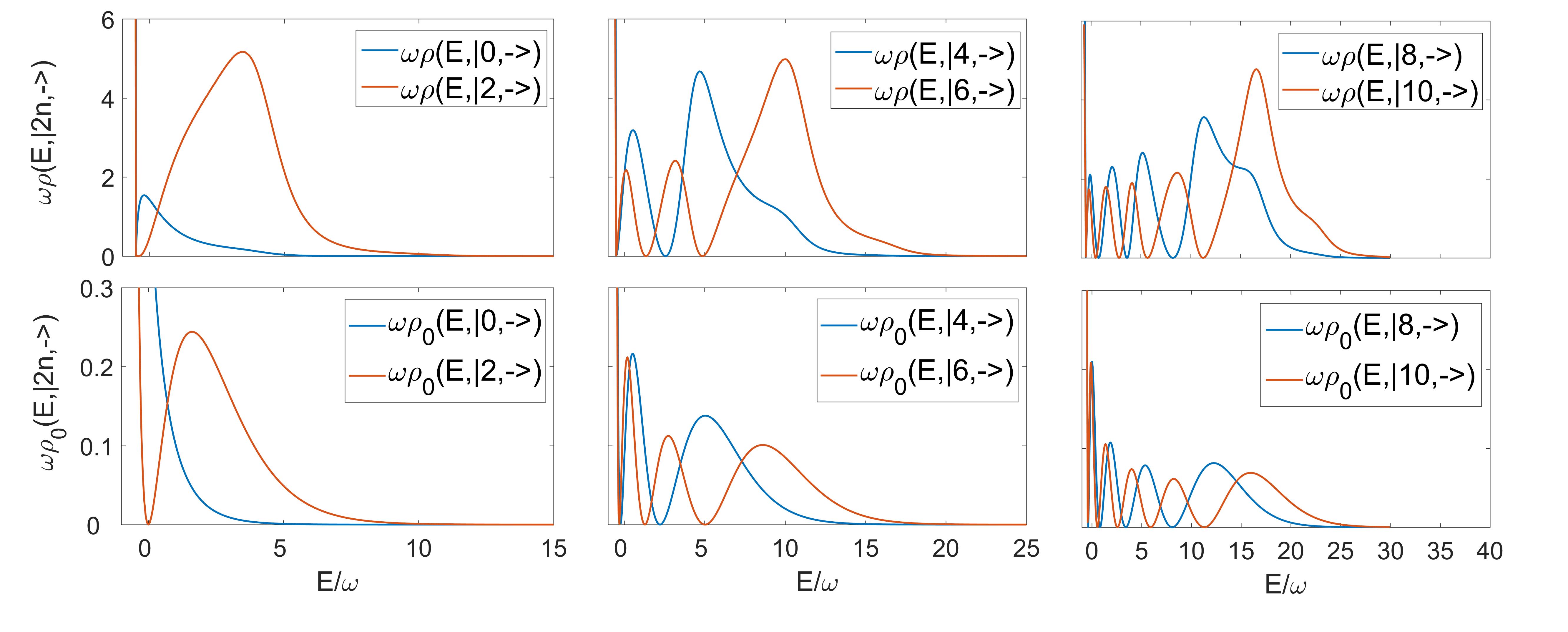}
\caption{\label{confrontia} Comparison between the the spectral functions related to the first six states belonging to $\tilde{S}_{+1}$, between the cases $\omega_0=0.8\omega$ and $\omega_0=0$ (whose analytic form is known) at the collapse point $g/\omega=0.5$. In all cases $\varepsilon=0.0005$, while the truncation number exploited for the continued fraction is  $N=24000$.}
\end{figure}

\begin{figure}[ht!]
\centering
\captionsetup{width=0.95\linewidth}
\includegraphics[height=5.8cm, width=14cm]{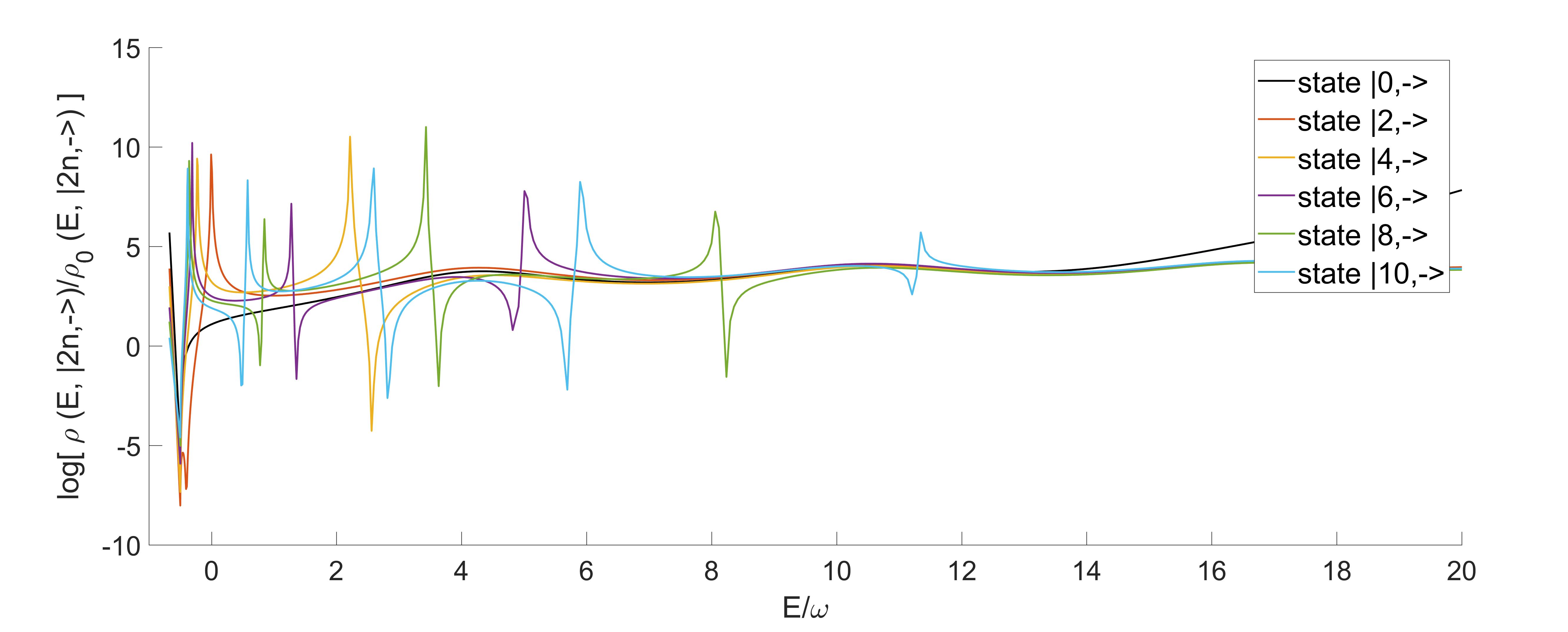}
\caption{\label{confrontib}Plot of the ratio between the two cases compared in figure \ref{confrontia}. It can be seen that for high energies the ratio between the spectral function in the case $\omega_0 \neq 0$ and the exact case $\omega_0=0$ is constant.}
\end{figure}

In this section we derive an expression of the spectral function of a factorised state $\ket{n, \sigma}$ in a continued fraction form.  The spectral function $\rho\left(E,\ket{\Psi},\ket{\Psi'}\right)$ is the matrix element $\rho(E)_{\Psi,\Psi'}$ of the microcanonical density operator defined in the following way:
\begin{equation}
\label{eq:defro}
\rho(E)= \delta(E-H)=\sum_\lambda \ket{\varepsilon_\lambda}\bra{\varepsilon_\lambda}\delta(E-E_\lambda)
\end{equation}
where $H$ is the Hamiltonian of the model and $\ket{\varepsilon_\lambda}$ is the eigenstate of $H$ related to the eigenvalue $E_\lambda$, $H\ket{\varepsilon_\lambda}=E_\lambda \ket{\varepsilon_\lambda}$. Other than in quantum statistical mechanics, it appears in relation with the resolvent $(E-H)^{-1}$, whose spectral representation is:
\begin{equation}
\label{eq:Rspec}
R_H(E) = \frac{1}{E-H} = \sum_\lambda \frac{\ket{\varepsilon_\lambda}\bra{\varepsilon_\lambda}}{E-E_\lambda} =\  \lim_{\epsilon\to 0^+} \int_{-\infty}^{+\infty} \frac{1}{E-E'-i\epsilon}\, \rho(E')dE'
\end{equation}
From the definition (\ref{eq:defro}) one can see that the diagonal element $\rho(E,\ket{\Psi})$ contains all the spectral information useful in the study of the state $\ket{\Psi}$ of the system. It can be in fact interpreted as the probability distribution of the state $\ket{\Psi}$ to be in a particular eigenstate of the Hamiltonian:
\begin{equation}
\label{12}
\rho(E,\ket{\Psi})= \sum_\lambda \left|\braket{\varepsilon_\lambda|\Psi} \right|^2 \delta\left(E-E_\lambda\right)\,,
\end{equation}
and it is instrumental in studying the time evolution of the state.\\
Its numerical computation can be rather involved if attacked in terms of Bargmann functions, though. Here we address this issue by making use of the connection of the spectral function to the resolvent of the system. The distributional identities $\lim_{\varepsilon \to 0} \frac{1}{x \pm i\varepsilon} = \mathrm{P}\frac{1}{x} \mp i\pi \delta (x)$, with $\varepsilon>0$ and  $\mathrm{P} $  principal part, determine
\begin{equation}
\label{ro}
\rho(E,\ket{\Psi}) = \frac{1}{\pi}\, \lim_{\varepsilon\to 0}\, \mathrm{Im}\braket{\Psi|\left(E-H-i\varepsilon\right)^{-1}|\Psi}\,.
\end{equation}
In the factorized states basis $\ket{n,\sigma}$, the resolvent of the QRM and 2$\gamma$QRM can readily be expressed in  continued fraction form (see \cite{Ziegler, BraakCF} or Appendix \ref{app1}), which makes a numerical calculation of the spectral function $\rho(E,\ket{n,\sigma})$ accessible. Notice that the use of continued fractions has been a staple in the treatment of the QRM, in different guises and forms \cite{Swain}. Taking the rotated Hamiltonian (\ref{eq:rotated}), the element of the resolvent related to the state $\ket{n,\sigma}$ is in the form:
\begin{equation}
\label{14}
\begin{split}
&{}\bra{n,\sigma}TR_{\tilde{H}}T^{\dagger}\ket{n,\sigma}\ =\\ 
&{}=\ \frac{1}{\left(z-A_{\big\lfloor \frac{n}{2} \big\rfloor}\right)\ \ -\ \ \cfrac{R_{{\big\lfloor \frac{n}{2} \big\rfloor}+1}^2}{\left(z-A_{{\big\lfloor \frac{n}{2} \big\rfloor}+1}\right)-\cfrac{R_{{\big\lfloor \frac{n}{2} \big\rfloor}+2}^2}{\quad\qquad\ddots\quad}}\ \ -\ \ \cfrac{R_{\big\lfloor \frac{n}{2} \big\rfloor}^2}{\left(z-A_{{\big\lfloor \frac{n}{2} \big\rfloor}-1}\right)-\cfrac{R_{{\big\lfloor \frac{n}{2} \big\rfloor}-1}^2}{\qquad\qquad\cfrac{\ddots\qquad}{z-A_0}}}}
\end{split}
\end{equation}
\vspace*{5pt}\\
where we set $z=E-i\varepsilon$ for the numerical calculation of (\ref{ro}) and the coefficients $A_j$ and $R_j$ depend on the subspace which $\ket{n,\sigma}$ belongs to: if the state has the form $\ket{2n,\pm(-1)^n}$ ($\,$i.e. it belongs to the subspace $S_{\mp1}\,$) the coefficients are $A_j= 2j\omega \pm (-1)^j\omega_0/2$ and $R_j=g\sqrt{2j(2j-1)}$; if the state is in the form $\ket{2n+1,\pm(-1)^n}$ ($\,$i.e. it belongs to the subspace $S_{\pm i}\,$) the coefficients are $A_j= (2j+1)\omega \mp (-1)^j\omega_0/2$ and $R_j=g\sqrt{2j(2j+1)}$.\\
Equation (\ref{14}) allows us to calculate the spectral function of any factorized state $\ket{n,\sigma}$ of the $2\gamma$QRM. In this work we show the results related to the positive parity subspace $S_{+1}$. Since we are working in the rotated basis, from now on we use $\ket{2n,-}$ as notation for the state belonging to $\tilde{S}_{+1}$.\\
The convergence of the continued fraction has been determined through Pringsheim's Theorem, under the condition that  $g<\omega/2$ (see Appendix \ref{app2}). 
The actual computation of the continued fraction expansion involves a truncation in Fock space for each truncation of the continued fraction.
\begin{figure}[t]
\center
\includegraphics[height=6.78cm, width=16cm]{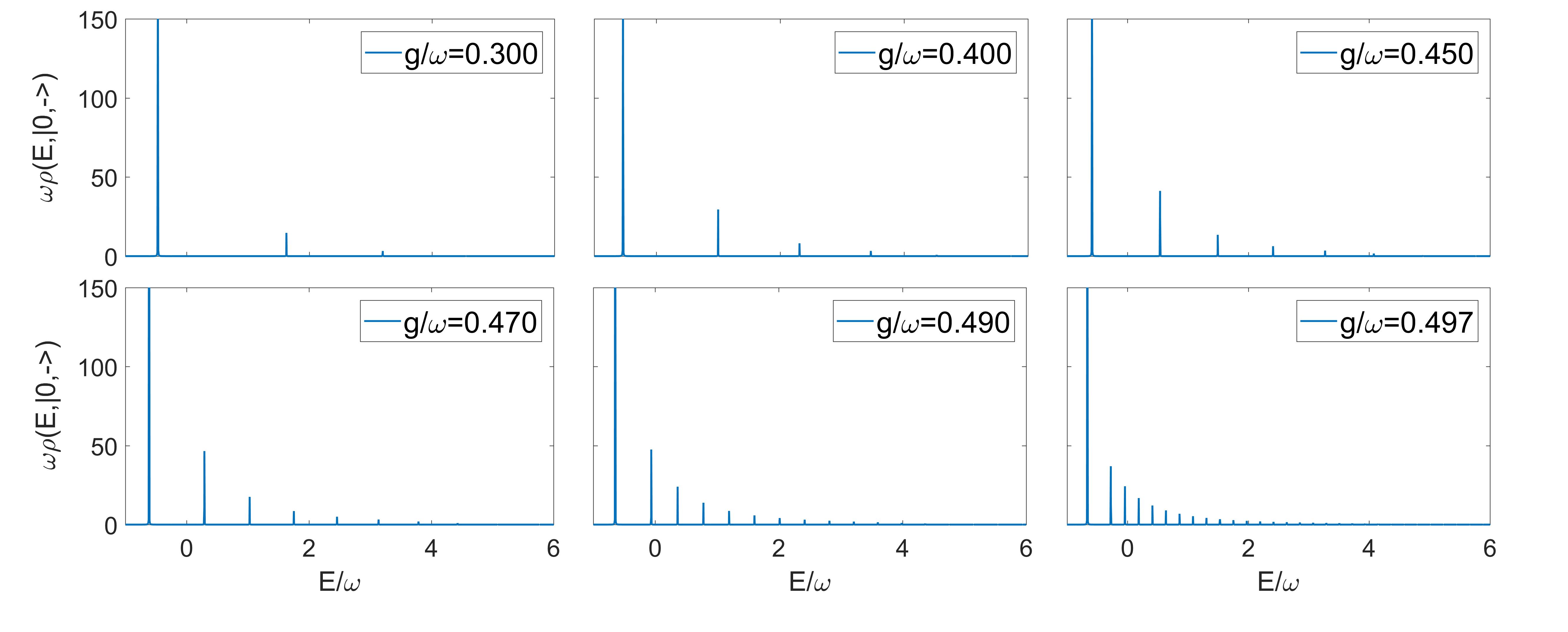}
\captionsetup{width=0.95\linewidth}
\caption{\label{fig:1} Spectral density related to the state $\ket{0,-}$ of the $2\gamma$QRM at different values of $g/\omega$. The value of the atomic frequency is $\omega_0=0.8\omega$ in all cases, while $\varepsilon=0.0005$ (see eq. (\ref{ro})). The truncation of the continued fraction is chosen in order to have convergence.}
\end{figure}
In figure \ref{fig:1} we report the numerical determination of the the spectral density for the vacuum state of the $2\gamma$QRM  at different values of $g/\omega$. Notice that the parameter $\epsilon$ has to be fixed for the numerical evaluation. Its value is chosen in such a way it does not affect the ratio between the peaks, and a smaller value would result only in a common scaling factor that does not bring improvement in the determination of the spectral function. \\
The method at hand, namely the numerical computation by continued fractions of spectral functions, allows us to insert much higher truncation numbers than with a direct simulation with truncation in Fock space, even  very close to the collapse point $g/\omega=0.5$, where the spectrum will no longer be purely discrete. Figure \ref{fig:2} shows the spectral density as we approach the special value $g/\omega=0.5$, making apparent this  change of the spectrum into an isolated discrete value and a continuum.\\
We now apply our technique to 
the collapse point $g/\omega=0.5$, even though  Pringsheim's theorem only guarantees convergence  in the discrete case $g/\omega < 0.5$. In fact the continued fraction approach allows only a discrete approximation of a continuum spectrum, but this is done at very high truncation numbers. In figure \ref{Om0a} the spectral function of the vacuum state for $g/\omega=0.5$ is calculated at different values of $\omega_0$. A first point of note is that the presence of an isolated ground state is linked to the atomic frequency $\omega_0$ being different from zero. Secondly, observe that the energy difference between the ground state and the continuum (figure \ref{Om0b}) is not linear in $\omega_0$, as observed also for $g/\omega < 0.5$ in previous papers \cite{Travenec,Travenec2, Duan}. \\
In the case $\omega_0=0$ the spectral function $\rho_0 \left(E, \ket{2n,-}\right)$, with $\ket{2n,-} \in \tilde{S}_{+1}$, can be calculated analytically. Consider the Hamiltonian of the 2$\gamma$QRM projected in $\tilde{S}_{+1}$ for a coupling value $g=\omega/2$:
\begin{equation}
\tilde{H}_+= \omega a^\dagger a + \frac{\omega}{2}\left( a^2 + (a^\dagger)^2\right) - \frac{\omega_0}{2} (-1)^{a^\dagger a / 2}
\end{equation}
We can express it in terms of $x$ and $p$ operators. In fact, knowing that $a=\sqrt{\frac{\omega}{2}}\left(x+ip/\omega\right)$ and $a^\dagger=\sqrt{\frac{\omega}{2}}\left(x-ip/\omega\right)$, we obtain:
\begin{equation}
\tilde{H}_+ = \omega^2 x^2 - \frac{\omega}{2} - \frac{\omega_0}{2} (-1)^{(x^2+p^2-1) / 4}
\end{equation}
In the case $\omega_0=0$ the Schr\"odinger equation takes the form $\left(\omega^2 x^2 - \omega/2\right)\ket{\Psi(x)}=E\ket{\Psi(x)}$ and the eigenstates coincide with the position operator eigenstates $\ket{x}$. Therefore, the spectral function related to the state $\ket{2n,-}$ can be expressed in terms of the Hermite polynomials $H_m(\xi)$:
\begin{equation}\label{eq:rho0hermite}
\rho_0 \left(E, \ket{2n,-}\right) = \frac{1}{4^n (2n)! \sqrt{\pi\,}}\, \frac{H^2_{2n}\big(\sqrt{E/\omega + 1/2\,}\,\big)}{ \sqrt{\omega\left(E + \omega/2\right)}}\, e^{-E/\omega - 1/2}
\end{equation}
These functions present a divergence at $E=-\omega/2$, while the zeros of $\rho_0(E,\ket{2n,-})$ are determined by the zeros of $\qquad$ $H_{2n}^2(\small{\sqrt{E/\omega+1/2}}\,)$. Notice further the normalization
\begin{equation}
  \label{eq:rho2n0normal}
  \int_{-\omega/2}^\infty \mathrm{d}E\,\rho_0\left(E,\ket{2n,-}\right)=1\,.
\end{equation}
We can now contrast and calibrate  the numerical results at $\omega_0\neq 0$ for the first six states of the subspace $\tilde{S}_{+1}$ 
with the corresponding analytical expression (\ref{eq:rho0hermite}), in figure \ref{confrontia}. Clearly the qualitative structure is well tracked by our numerical procedure, setting aside the divergence of $\rho_0$ at $E=-\omega/2$. In particular, notice the number of nodes in the corresponding spectral functions. Moreover, in figure \ref{confrontib} we plot the ratio between the two quantities. Even if the polynomial trend of the truncated continued fraction can not track the exponential trend of (\ref{eq:rho0hermite}), in a range of high energies for which the Hermite trend contributes mostly, we can notice a constant value which is due to the atomic term in the Hamiltonian (\ref{9}) becoming progressively less relevant. 
\newpage

\section{The Survival Probability of the vacuum state} \label{sec4}
The results of the previous section can be exploited for the determination of an important dynamical quantity: the survival probability, that is,  the probability of finding the system in its initial state after a time evolution of interval $t$.\\
The connection between the spectral function $\rho \left(E, \ket{\Psi}\right)$ and the survival probability is given through the survival amplitude $A_\Psi(t)=\braket{\Psi|U(t)|\Psi}=\braket{\Psi|e^{-iHt}|\Psi}$  by Fourier transform,
\begin{equation}
\int \mathrm{d}E\, e^{-iEt} \rho\left(E,\ket{\Psi}\right) =  \int \mathrm{d}E\, e^{-iEt} \sum_\lambda \left|\braket{\varepsilon_\lambda|\Psi} \right|^2 \delta\left(E-E_\lambda\right)= \sum_\lambda \left|\braket{\varepsilon_\lambda|\Psi} \right|^2 e^{-iE_\lambda t} =  \braket{\Psi|e^{-iHt}|\Psi}\,.
\end{equation}
That is,
\begin{equation}
P_\Psi(t) = \left|A_\Psi(t) \right|^2 = \left| \int dE e^{-iEt} \rho\left(E,\ket{\Psi}\right)\right|^2
\end{equation}
with the integration on the domain defined by $\rho(E,\ket{\Psi})$.\\
Let us now focus  on the vacuum state of the 2$\gamma$QRM. It is of interest since it can be prepared as  the ground state in the decoupled or strong coupling regime ($g/\omega \ll 0.1$), and then adiabatically moved to larger couplings. In terms of the eigenenergies $E_\lambda$ and the transition probabilities $\left|\braket{\varepsilon_\lambda|0,-}\right|^2$, which can be derived from its spectral function, the survival probability of the vacuum state $\ket{0,-}$ can be written as:
\begin{equation}
\label{21}
P_{\ket{0,-}}(t) = \left|\sum_\lambda \left| \braket{\varepsilon_\lambda|0,-}\right|^2 e^{-iE_\lambda t} \right|^2
\end{equation}
This connection provides us with a numerical technique to compute the survival probability, through numerical computation of the spectral function. The fact that we do not use matrix inversion,  diagonalization, nor exponentiation in the process means that the point of the truncation can be much higher than what could be reasonably achieved with Fock space expansions for the survival probability. This numerical advantage allows us, in particular, an analysis of the survival probability for  the 2$\gamma$QRM close to the collapse point $g=\omega/2$.\\
\begin{figure}[t]
\captionsetup{width=.95\linewidth}
\qquad\includegraphics[height=6.78cm, width=16cm]{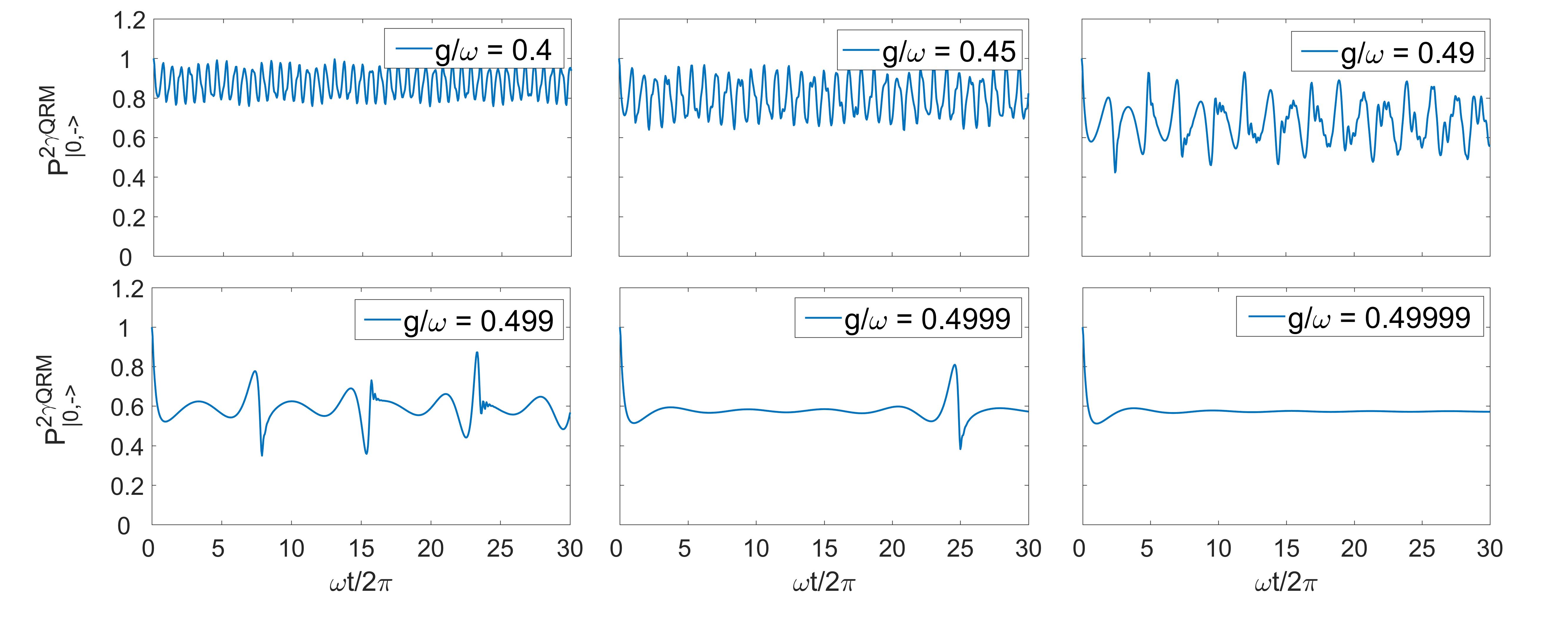}
\caption{\label{SPa} Survival Probability related to the vacuum state $\ket{0,-}$ of the 2$\gamma$QRM, approaching to the collapse point $g/\omega=0.5$. The quantity is calculated through the spectral function of the state considered (see section \ref{sec4}). For all cases $\omega_0=0.8\omega$.}
\end{figure}
\begin{figure}
\centering
\captionsetup{width=0.95\linewidth}
\includegraphics[height=6.78cm, width=16cm]{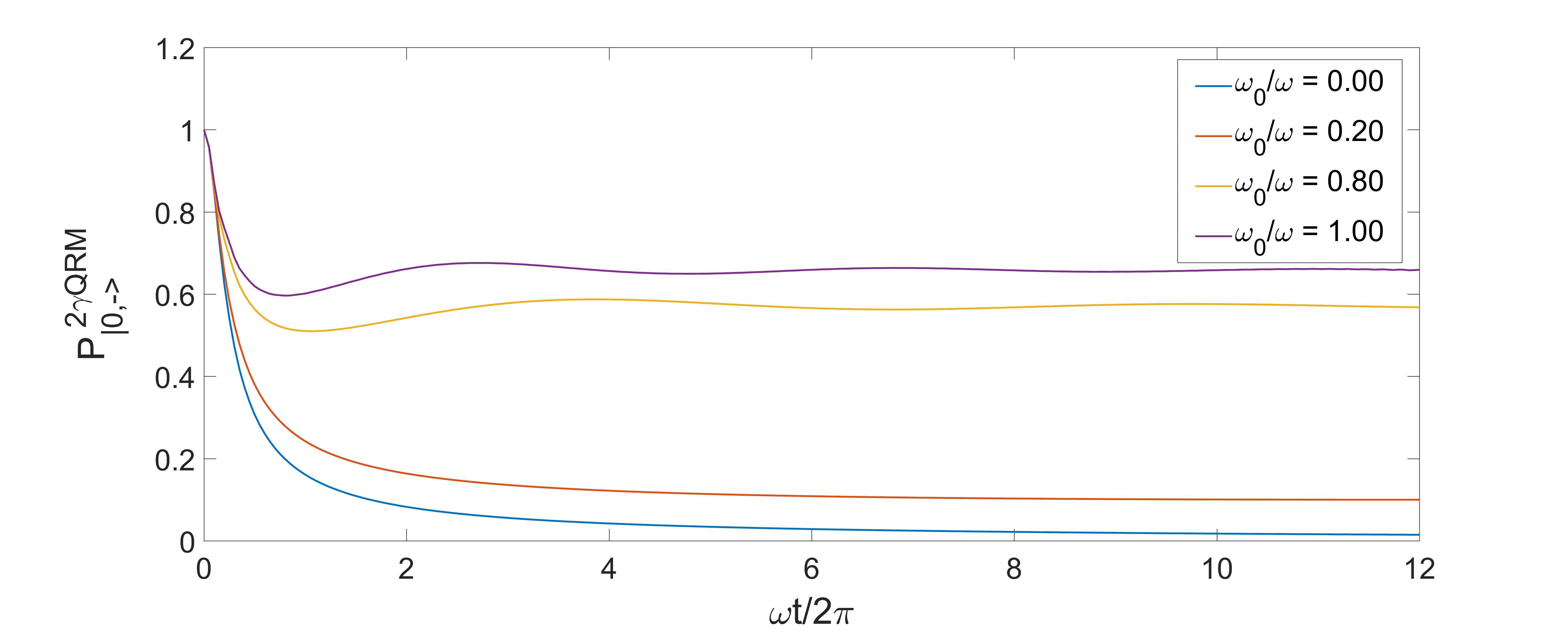}
\caption{\label{SPb} Survival Probability of the vacuum state $\ket{0,-}$ of the 2$\gamma$QRM in correspondence of the collapse point of the spectrum $g/\omega=0.5$, at different values of the atomic frequency $\omega_0$. The case $\omega_0=0$ is consistent with the analytical result (\ref{23}).}
\end{figure}
In figure \ref{SPa} we report the numerical determination of the survival probability at different values of $g/\omega$. Near the collapse point $g=\omega/2$  interference effects become predominant. This was to be expected from the spectral density depicted in figure \ref{fig:2}, since the  density of eigenstates means that small frequencies (small energy differences) will play a major role in the survival probability. Indeed the long time behaviour of the survival probability becomes flatter, as seen in the last graph of figure \ref{SPa}.\\
We also compute the survival probability for $g/\omega=0.5$ at different values of $\omega_0$, as portrayed in figure \ref{SPb}. We again see that the survival probability for $\ket{0,-}$ presents a dominant constant value, dependent on the atomic parameter, after a short transient.  This can be understood by looking at the form of the Survival Probability $P_{\ket{0,-}}(t)$ in terms of the spectral function,
\begin{equation}
  P_{\ket{0,-}}(t)= \left|\ \left|\braket{\varepsilon_0|0,-}\right|^2 + \int_{-\omega/2}^{\infty}\mathrm{d}E\,\rho\left(E,\ket{0,-}\right)e^{-iEt} \right|^2
\end{equation}
and application of the Riemann--Lebesgue lemma. Indeed, we know that $\rho\left(E,\ket{0,-}\right)$ is integrable - in fact, as pointed out above, it is normalized to $1$. Therefore the Fourier transform above tends to zero at infinity. To be more precise, only the discrete part of the spectrum contributes to the long time behaviour,
\begin{equation}
  \label{eq:limitsurvival}
\lim_{t\to\infty}P_{\ket{0,-}}(t)  =  \left|\braket{\varepsilon_0|0,-}\right|^4\,.
\end{equation}
Moreover, the case $\omega_0=0$ (the blue line in figure \ref{SPb}) agrees with the analytical exact result from $\rho_0 \left( E, \ket{0,-} \right)$:
\begin{equation}
\label{23}
P_{\ket{0,-}}(t)=\left| \int \mathrm{d}E\, e^{-iEt} \rho_0 \left( E,\ket{0,-}\right) \right|^2 =\left| \frac{e^{i\omega t/2}}{\sqrt{\pi \omega}} \int_{-\omega/2}^{\infty} \mathrm{d}E\, \frac{e^{-(1+i\omega t)(E/\omega+1/2)}}{\sqrt{E/\omega+1/2}}\right|^2 = \frac{1}{\pi } \Big| \int_{-\infty}^{+\infty} \mathrm{d}\epsilon\, e^{-(1+i\omega t)\epsilon^{2}} \Big|^2 = \frac{1}{\sqrt{1+\omega^2 t^2}}\,.
\end{equation}
Notice the asymptotic $1/t$ behaviour, that is due to the $1/(E+\omega/2)^{1/2}$ divergence in the integrand.\\
As $\omega_0$ grows, a discrete point will appear in the spectrum, and thus a constant term in the long time behaviour of the survival probability. The subleading term will be generically of form $1/t$, since the leading behaviour of the Fourier transform of the continuum part will be $1/t$ or faster decay.

\section{Conclusions and Perspectives} \label{conclusions}
In this work we have studied numerically spectral functions for the Two Photon Quantum Rabi Model (2$\gamma$QRM) and the corresponding survival probabilities. These two quantities are more readily amenable to numerical treatment than direct diagonalization of the Hamiltonian, as is shown by the much higher truncation numbers we can achieve in this approach.\\
Since there are indeed several proposals for quantum simulation implementation of the 2$\gamma$QRM \cite{Felicetti,PhysRevA.95.063844,SolanoSimSC,Schneeweiss:2017aa}, our improved numerical approach will prove beneficial for their analysis.\\
This improvement of numerics has allowed us to investigate further the collapse point, at which the spectrum becomes continuous. This is indeed the result recovered both from spectral functions and from survival probabilities.\\
As all numerics are suspect in the environment of a drastic structural change, such as the spectral collapse at hand, we have proposed an independent check by comparing spectral functions at the collapse point for the exactly solvable case with $\omega_0=0$, expressed in terms of Hermite polynomials, with those corresponding to $\omega_0\neq0$. The qualitative structure, in particular the number of modes and the large energy/short time behaviours, is maintained as expected, thus providing us with a calibration tool.\\
In particular we note that a signature of the collapse of the spectrum into a purely continuous one would be that all survival probabilities necessarily tend to zero. In the case at hand there is a remaining relevant discrete point in the spectrum, and the long time limit of the survival probability is a constant, determined by the projection of the initial state onto the corresponding proper eigenstate.\\
The direct measurement of such a phenomenon in the survival probability might not be immediately possible in the different platforms in which the 2$\gamma$QRM is a good description of the dynamics for some range of the parameters. However, there are alternatives to detect the spectral collapse, one of which we now put forward. Consider thus that there is another eigenstate of the full system which can be coupled to the discrete element of the $\tilde{S}_{+1}$ subspace. In such a situation, the long term behaviour of the survival probability for any state in the $\tilde{S}_{+1}$ subspace will be given by coherent Rabi oscillations, providing us with a target for detection. Notice a very recent alternative proposal to investigate the spectral collapse, in this case for the 2$\gamma$QRM with full quadratic coupling, studying a two-time correlation for the output field in a driven system \cite{Felicetti:2018aa}. \\
In summary, we have investigated further the rich phenomenology of the 2$\gamma$QRM, with emphasis on the numerically computable spectral functions and survival probability, and we suggest new avenues for the exploration of the spectral collapse.

\begin{appendices}

\section{The continued fraction form of the resolvent} \label{app1}
In each subspace of defined four-parity $\Pi_4$  the rotated Hamiltonian $\tilde{H}$ is  tridiagonal  in the basis of Fock states. For instance, the 2$\gamma$QRM Hamiltonian projected in the subspace of positive parity $\tilde{S}_{+1}$ is (see eq. (\ref{9})):
\begin{equation}
\tilde{H}_{+1} = \omega a^\dagger a + g\left(a^2 + (a^\dagger)^2\right) - \frac{1}{2}\omega_0 \cos\Big(\frac{a^\dagger a}{2} \pi\Big)
\end{equation}
Since the elements $\braket{2n,-|\tilde{H}_{+1}|2m,-}$ are non-zero only if $m=n, n\pm1$, the matrix form of the projected Hamiltonian assumes a tridiagonal form in the basis $\ket{2n,-}$:\\
\begin{equation}
\label{tridiagH}
\tilde{H}_{+1} =
\begin{pmatrix}
A_0 & R_1 & 0 & 0 & \cdots &  \\ 
R_1 & A_1 & R_2 & 0 & \cdots &  \\ 
0 & R_2 & A_2 &  R_3 & \qquad & \\
\vdots & \quad & \ddots & \ddots & \ddots   \\
\end{pmatrix}
\end{equation}
where $A_n= \braket{2n|\tilde{H}_{+1}|2n}= 2n \omega - (-1)^n \omega_0 /2$ and $R_n=\braket{2n|\tilde{H}_{+1}|2n-2}=\braket{2n-2|\tilde{H}_{+1}|2n}= g\sqrt{2n(2n-1)}$.\\
As regards to the resolvent related to this Hamiltonian, $R_{\tilde{H}_{+1}}(z)= (z-\tilde{H}_{+1})^{-1}$, we can see that its diagonal elements $\braket{2n,-|R_{\tilde{H}_{+1}}(z)|2n,-}$ can be expressed in a continued fraction form. If one is interested only in the first element $\braket{0,-|R_{\tilde{H}_{+1}}(z)|0,-}$ the continued fraction form can be achieved also through the Recursive Projection Method \cite{Ziegler}. However, since it is a general property of the tridiagonal matrices one can exploit the iterative relation of tridiagonal matrix minors for the determination of a generic element of the resolvent \cite{BraakCF}. \\
In this section we show how to obtain equation (\ref{14}), namely the continued fraction form of the element $\braket{n, \sigma|TR_{\tilde{H}}T^{\dagger}|n,\sigma}$, with $\ket{\Psi}=T^{\dagger}\ket{n, \sigma}$ state belonging to any one of the four subspaces (\ref{subspaces}). We derive the expression for $\ket{\Psi}\equiv \ket{2n,-} \in \tilde{S}_{+1}$. Since the form of the Hamiltonian $\tilde{H}$ in any of the subspaces $\tilde{S}_{\pm 1, \pm i}$ is the same as (\ref{tridiagH}), the derivation of the diagonal element of the resolvent does not change from the one in the subspace $\tilde{S}_{+1}$.\\
From the theory of linear algebra the inverse of a square matrix $Q$ is the matrix of elements $Q_{ij}=\Delta_{ij}/\mathrm{det}(Q)$, where $\Delta_{ij}=(-1)^{i+j}\mathrm{det}(M_{ij})$ is the $(i,j)$-cofactor and $M_{ij}$ is the first minor, obtained by eliminating the $i$-th row and the $j$-th column. We use the following notation: $D_0$ as the determinant of $\big(z-\tilde{H}_{+1}\big)$, $D_k$ as the corresponding determinant of the matrix resulting from eliminating the first $k$ rows and $k$ columns,  and $\tilde{D}_k$ as the determinant of the matrix given by the restriction to  the first $k+1$ rows and columns of  the same matrix. So we have:
\begin{eqnlist}
&{} D_0=\ \mathrm{det}\Big(z-\tilde{H}_{+1}\Big) \\ &{}
D_n = \mathrm{det}  
\begin{pmatrix}
z-A_n		&  R_{n+1}    & 0             & \cdots \\
R_{n+1}	& z-A_{n+1}   & R_{n+2}  & \cdots \vspace*{3pt} \\ 
	\vdots	&	\qquad\qquad\ddots &  &\\ \\
\end{pmatrix} \quad
\tilde{D}_n = \mathrm{det} 
\begin{pmatrix}
z-A_0		&  R_1	    & \cdots &\\
R_1		& z-A_1	& R_2 &\vspace*{3pt}\\ 
				&	 & \ddots\	& R_{n} \vspace*{3pt}\\
	& \cdots  & R_{n}    & z-A_{n} \\
\end{pmatrix} 
\end{eqnlist}
Since the sub-matrix obtained by removing the $n$-th row and $n$-th column of $(z-\tilde{H}_{+1}\big)$ has a block-tridiagonal form, applying the formula for the inverse matrix we have:
\begin{equation}
\label{27}
\braket{2n,-|\big(z-\tilde{H}_{+1}\big)^{-1}|2n,-}= \frac{\Delta_{nn}}{\mathrm{det}(z-\tilde{H}_{+1}\big)} = \frac{\tilde{D}_{n-1}D_{n+1}}{D_0}
\end{equation}
Moreover, $D_0$ can be written in terms of $\tilde{D}_{n-1}$ and $D_{n+1}$ using the Laplace formula for the matrix determinant:
\begin{equation}
\label{28}
 D_0= \mathrm{det}(z-\tilde{H}_{+1}\big) = (z-A_n)\tilde{D}_{n-1}D_{n+1} - R_{n+1}^2 \tilde{D}_{n-1}D_{n+2} - R_{n}^2 \tilde{D}_{n-2}D_{n+1}
\end{equation}
Thus eq. (\ref{27}) takes the form:
\begin{equation}
\frac{\tilde{D}_{n-1}D_{n+1}}{D_0}= \frac{1}{(z-A_n) - R_{n+1}^2\, D_{n+2}/D_{n+1} - R_{n}^2\, \tilde{D}_{n-2}/\tilde{D}_{n-1}}
\end{equation}
Using similar arguments a recursive formula for $D_{n}/D_{n-1}$ and $\tilde{D}_{n}/\tilde{D}_{n+1}$ can be obtained:
\begin{equation} 
\label{29}
\frac{D_{n}}{D_{n-1}}= \frac{1}{(z-A_{n-1}) - R_n^2 \, D_{n+1}/D_{n}} \qquad
\frac{\tilde{D}_{n}}{\tilde{D}_{n+1}}= \frac{1}{(z-A_{n+1}) - R_{n+1}^2 \,\tilde{D}_{n-1}/\tilde{D}_{n}}
\end{equation}
Applying iteratively the two formulas (\ref{29}) the two quantities $D_{n}/D_{n-1}$ and $\tilde{D}_{n}/\tilde{D}_{n+1}$ can be expressed in a continued fraction and a finite continued fraction form respectively:
\begin{eqnlist} 
&{} D_{n}/D_{n-1} = \frac{1\qquad}{\ (z-A_{n-1})\ -}\ \ \frac{R_{n}^2\qquad }{\ (z-A_n)\ -}\ \ \frac{R_{n+1}^2\qquad }{\ (z-A_{n+1})\ -} \cdots \\
&{} \tilde{D}_{n}/\tilde{D}_{n+1} = \frac{1\qquad}{\ (z-A_{n+1})\ -}\ \ \frac{R_{n+1}^2\quad }{\ (z-A_n)\ -}\cdots \frac{\ R_1^2\qquad }{\ (z-A_0)}
\end{eqnlist}
Substituting them in equation (\ref{29}) we obtain the continued fraction form of a diagonal element of the resolvent (\ref{14}).

\vspace*{10pt}

\section{Convergence of the continued fraction expansion} \label{app2}
We want to study the convergence of the continued fraction form we have given for the element $\braket{0,-|R_{\tilde{H}_{+1}}|0,-}$ of the resolvent:
\begin{equation}
\braket{0,-|R_{\tilde{H}_{+1}}|0,-}= \frac{1}{\ E-i\varepsilon + \omega_0/2\ -}\ \ \frac{2g^2}{\ E-i\varepsilon-2\omega- \omega_0/2\ -}\ \ \frac{12g^2}{\ E-i\varepsilon-4\omega+ \omega_0/2\ -}\cdots
\end{equation}
Define for this  case $a_0=1$,  $a_n=g^2 2n(2n-1)$ for $n\geq1$, and $b_n=E-i\epsilon-2n\omega+(-1)^n\omega_0/2$ for $n\geq0$. The continued fraction at hand is
\begin{equation}
\braket{0,-|R_{\tilde{H}_{+1}}|0,-}= \frac{a_0\ }{b_0\ -}\ \ \frac{a_1\ }{\ b_1\ -}\ \ \frac{a_2\ }{\ b_2\ -} \cdots\,.
\end{equation}
This continued fraction can be written as
\begin{equation}
   \frac{1/\omega}{\ \alpha_0\ -}\ \ \frac{1}{\ \alpha_{1}\ -} \ \ \frac{1}{\ \alpha_{2}\ -} \dots \, .
\label{eq:standardcont}
\end{equation}
with
\begin{equation*}
  \alpha_{n} = c_nb_{n}\,,\qquad
  c_{n+1} = \frac{1}{a_{n+1}c_n}\,,\qquad
  c_0 = 1/\omega\,.
\end{equation*}
 The form of $a_n$ makes it easy to provide an explicit form, namely
\begin{eqnarray*}
  \label{eq:alphas}
  \alpha_{2n+1}&=& \frac{\omega b_{2n+1}}{a_{2n+1}} \prod_{l=1}^n \frac{a_{2l}}{a_{2l-1}} = \frac{\omega b_{2n+1}}{a_{2n+1}} \frac{\Gamma\left(\frac{1}{2}\right)\Gamma\left(\frac{1}{4}\right)}{\Gamma\left(\frac{3}{4}\right)} \frac{\Gamma\left(n+1\right)\Gamma\left(n+\frac{3}{4}\right)}{\Gamma\left(n+\frac{1}{2}\right)\Gamma\left(n+\frac{1}{4}\right)}\,.\\
  \alpha_{2n} &=&  \frac{b_{2n}}{\omega}\prod_{l=1}^n \frac{a_{2l-1}}{a_{2l}} = \frac{b_{2n}}{\omega}\frac{\Gamma\left(\frac{3}{4}\right)}{\Gamma\left(\frac{1}{2}\right)\Gamma\left(\frac{1}{4}\right)} \frac{\Gamma\left(n+\frac{1}{2}\right)\Gamma\left(n+\frac{1}{4}\right)}{\Gamma\left(n+1\right)\Gamma\left(n+\frac{3}{4}\right)} \,.
\end{eqnarray*}
Notice that the coefficients $\alpha_n$ are adimensional.
We can now use the Stirling approximation to obtain their asymptotic behaviour, 
\begin{eqnarray*} \alpha_{2n+1}&=&-\frac{4\omega^2}{16g^2}\frac{\Gamma\left(\frac{1}{2}\right)\Gamma\left(\frac{1}{4}\right)}{\Gamma\left(\frac{3}{4}\right)}+O\left(n^{-1}\right)\,,\\
  \alpha_{2n}&=& -4 \frac{\Gamma\left(\frac{3}{4}\right)}{\Gamma\left(\frac{1}{2}\right)\Gamma\left(\frac{1}{4}\right)}+O\left(n^{-1}\right)\,.
\end{eqnarray*}
We shall now see that, if $g^2<\omega^2/4$, Pringsheim's sufficient convergence criterion allows us to conclude convergence for complex energy. In order to see this, consider rewriting the continued fraction as 
\begin{equation}
   \frac{1/\omega}{\ \alpha_0\ -}\ \ \frac{1}{\ \alpha_{1}\ -}\ \ \frac{1}{\ \alpha_{2}\ -} \cdots\,=\frac{c/\omega}{\ \beta_0\ -}\ \ \frac{1}{\ \beta_{1}\ -} \ \ \frac{1}{\ \beta_{2}\ -}\cdots\, ,
\label{eq:standardcontbeta}
\end{equation}
for some real $c$. This is achieved by defining the coefficients $\beta_n$ by $\beta_{2n+1} = \frac{\alpha_{2n+1}}{c}$ and $
  \beta_{2n} = c \alpha_{2n}$.
Let $0<\delta\ll 1$ be a small positive real number. Choose $c$ as
\begin{equation}
  \label{eq:cvalue}
  c=- \frac{1}{2+\delta}\frac{\omega^2}{4 g^2}\frac{\Gamma\left(\frac{1}{2}\right)\Gamma\left(\frac{1}{4}\right)}{\Gamma\left(\frac{3}{4}\right)}\,.
\end{equation}
Then, asymptotically,
\begin{eqnarray*}
  &&\beta_{2n+1} = 2+\delta+O\left(n^{-1}\right)\,,\\
  &&\beta_{2n} = \frac{1}{2+\delta}\frac{\omega^2}{g^2}+O\left(n^{-1}\right)\,.
\end{eqnarray*}
In order to fulfill, asymptotically, Pringsheim's criterion $\left|\beta_n\right|>2$, we require
\begin{equation*}
   \frac{1}{2+\delta}\frac{\omega^2}{g^2} >2\,,
 \end{equation*}
 whence
 \begin{equation*}
   g^2 < \frac{\omega^2}{2(2+\delta)}\ ,
 \end{equation*}
 thus inside the normal region before the collapse.\\
The asymptotic convergence does not guarantee convergence of the resolvent if the energy is  one of the real eigenvalues; it is enough for our purposes, though, since in order to determine the relevant spectral function we have to compute the limit of the imaginary part of the resolvent.\\ \\

\end{appendices}

\section*{Acknowledgments}
E.L. acknowledges fruitful discussions with D. Braak. I.L.E. and E.S. acknowledge funding from Spanish MINECO/FEDER FIS2015-69983-P and the Basque Government IT986-16.

\section*{Author Contributions}
E.L. has performed the main calculations and numerical simulations and has been responsible together with I.E. for the writing of the paper. I.E. has contributed to theoretical calculations. A.M., A.N. and E.S. have contributed to the generation and development of the ideas and supervise the project throughout all stages.  

\section*{Additional Information}
\paragraph{Competing interests:}
The authors declare that they have neither financial nor non-financial competing interests.
\end{document}